\documentclass[12pt]{article}
\usepackage{epsfig, amsmath, amssymb}
\usepackage{graphicx,psfrag}
\usepackage{xcolor}
\newcommand{\be}{\begin{equation}}
\newcommand{\ee}{\end{equation}}
\newcommand{\ben}{\begin{equation}}
\newcommand{\een}{\end{equation}}
\newcommand{\bea}{\begin{eqnarray}}
\newcommand{\eea}{\end{eqnarray}}
\newcommand{\bA}{\begin{array}}
\newcommand{\eA}{\end{array}}
\newcommand{\bc}{\begin{center}}
\newcommand{\ec}{\end{center}}

\newcommand{\ra}{\rightarrow}
\newcommand{\del}{\partial}

\newcommand{\ie}{{\it i.e.}}
\newcommand{\eg}{{\it e.g.}}

\newcommand{\lan}{\langle}
\newcommand{\ran}{\rangle}

\begin{document}


\begin{titlepage}

%

\bc

\hfill 
\\         [25mm]

{\Huge On de Sitter future-past extremal surfaces
 \\ [2mm]  and the ``entanglement wedge''} 
\vspace{16mm}

{\large K.~Narayan} \\
\vspace{3mm}
{\small \it Chennai Mathematical Institute, \\}
{\small \it H1 SIPCOT IT Park, Siruseri 603103, India.\\}

\ec
\vspace{30mm}

\begin{abstract}
  We develop further the codim-2 future-past extremal surfaces
  stretching between the future and past boundaries in de Sitter
  space, discussed in previous work. We first make more elaborate the
  construction of such surfaces anchored at more general subregions of
  the future boundary, and stretching to equivalent subregions at the
  past boundary. These top-bottom symmetric future-past extremal
  surfaces cannot penetrate beyond a certain limiting surface in the
  Northern/Southern diamond regions: the boundary subregions become
  the whole boundary for this limiting surface. For multiple disjoint
  subregions, this construction leads to mutual information vanishing
  and strong subadditivity being saturated.  We then discuss an
  effective codim-1 envelope surface arising from these codim-2
  surfaces. This leads to analogs of the entanglement wedge and
  subregion duality for these future-past extremal surfaces in de
  Sitter space.
\end{abstract}


\end{titlepage}

{\tiny 
\begin{tableofcontents}
\end{tableofcontents}
}



\section{Introduction}

de Sitter space is of great interest for various reasons:
theoretically there is the striking fact that it has thermodynamic
properties, with temperature and entropy \cite{Gibbons:1977mu}\ (see
the review \cite{Spradlin:2001pw}). This entropy arises as the area of the
cosmological horizon in the static patch coordinatization for
observers in the Northern/Southern diamond regions who view these as
event horizons. It is fascinating to ask how de Sitter entropy can be
understood via gauge/gravity duality
\cite{Maldacena:1997re,Gubser:1998bc,Witten:1998qj, Aharony:1999ti}
for de Sitter space, or $dS/CFT$
\cite{Strominger:2001pn,Witten:2001kn,Maldacena:2002vr}, which
associates a hypothetical non-unitary dual Euclidean CFT at the future
boundary $I^+$, which might be regarded as the natural boundary of de
Sitter space (see \eg\ \cite{wittenStrings98}). The $dS/CFT$ dictionary
$\Psi_{dS}=Z_{CFT}$\ \cite{Maldacena:2002vr}, with $\Psi_{dS}$ is the
late-time Hartle-Hawking Wavefunction of the Universe with appropriate
boundary conditions and $Z_{CFT}$ the dual CFT partition function, is
quite different from $Z_{bulk}=Z_{CFT}$ in the $AdS$ case. For instance,
for $dS_4$, we have (semiclassically)
\bea\label{dS4/CFT3}
&& \qquad\qquad\qquad\quad Z_{{}_{CFT}}=\Psi_{_{dS}} \sim e^{iS_{cl}}\sim 
e^{-\int_k R_{dS}^2 k^3 \varphi_{-k}^0\varphi_k^0\,+\,\ldots}\ , \nonumber\\
&&  \lan O_kO_{k'}\ran\sim
{\delta^2Z_{{}_{CFT}}\over\delta\varphi_k^0\delta\varphi_{k'}^0}\ , \qquad\quad
\lan\varphi_k\,\varphi_{k'}\ran \sim \int D\varphi\ \varphi_k\,\varphi_{k'}\,
\big|\Psi_{dS}[\varphi_k]\big|^2\ .
\eea
The $CFT_d$ energy momentum tensor 2-point correlation functions yield
central charges that are negative or imaginary (odd dimensions),
effectively analytic continuations from $AdS$: \eg\ taking $O_k$ as
appropriate $T_{ij}$ components gives the real, negative, central charge 
$-{R_{dS}^2\over G_4}$\, reminiscent of ghost-like ($c<0$) theories.
In \cite{Anninos:2011ui}, a higher spin $dS_4$ duality was
conjectured involving a 3-dim CFT of anti-commuting (ghost) scalars,
which exemplifies this (see also \eg\
\cite{Bousso:2001mw}-\cite{Anninos:2017eib}).\
While dual operator correlation functions are obtained by a differentiate
prescription applied to $Z_{CFT}$, bulk expectation values are obtained
by weighting with the bulk probability $|\Psi_{dS}|^2$. 
The fact that bulk observables in this formulation require both $\Psi_{dS}$
and $\Psi_{dS}^*$ suggests that two copies of the dual CFT are required
for a fixed $dS$ background (strictly one should also sum over final
3-metrics in $|\Psi_{dS}[\varphi,g^3]|^2$).

In this context it is interesting to ask if the various ideas and
techniques pertaining to holographic entanglement unravelled in
$AdS/CFT$ \cite{Ryu:2006bv,Ryu:2006ef,HRT} (reviewed in
\eg\ \cite{Rangamani:2016dms,Harlow:2018fse,Headrick:2019eth}) have
analogs in de Sitter space, perhaps leading to insights into de Sitter
entropy as some sort of generalized entanglement entropy. In the $AdS$
case, the areas of extremal surfaces anchored at the boundary of
subsystems in the boundary theory encode the entanglement entropy of
the subsystem in the dual field theory. It is known that $AdS$ is
special in many ways: many apparently gravitational or geometric
quantities are actually field theory quantities. For instance, the
extremal surfaces encoding entanglement are geometric objects but very
strikingly they automatically satisfy the various inequalities that
entanglement entropy in field theory is required to satisfy.  In these
cases, the dual field theory includes the time direction. It is thus
not clear if any of these ideas and mathematical formulations of
entanglement make sense away from $AdS/CFT$, or more generally
gauge/gravity duality for ordinary field theories. In de Sitter, the
natural boundary is at (future or past) timelike infinity and is
spatial so the dual is hypothesised to be a Euclidean nonunitary CFT.

One way to set up the analog of the Ryu-Takayanagi formulation in
$dS$, for one thing simply as a geometric problem, is to look for
extremal surfaces pertaining to subregions at the future
boundary\ (see \cite{Narayan:2019pjl} for a review of these
investigations).  Since the theory is Euclidean, there is no natural
time direction: as a calculational crutch, one could pick one of the
spatial symmetry directions as boundary Euclidean time and look for
extremal surfaces on bulk slices corresponding to these. All such
slices must be equivalent however since none of these is
sacrosanct. In the Poincare slicing, this exercise shows that surfaces
that begin at $I^{+}$ do not turn back somewhere in the bulk to return
to $I^{+}$: there is no real turning point for such timelike surfaces
ending on the spatial boundary. There are complex extremal surfaces
with turning points, amounting to analytic continuation from the $AdS$
Ryu-Takayanagi surfaces but their interpretation is unclear.

In \cite{Narayan:2017xca}, certain codim-2 timeline extremal surfaces 
were found stretching from the future boundary to the past: this is 
perhaps natural given that surfaces do not return to $I^{+}$, so they 
could instead end at $I^{-}$. This also dovetails with the fact that 
bulk expectation values require two copies of the wavefunction and 
so two CFT copies and therefore two boundaries. These surfaces 
begin at $I^{+}$, the future boundary of the future universe $F$, 
have a turning point in the Northern/Southern diamond regions $N/S$ 
and then end at the past boundary $I^{-}$ of the past universe $P$.
These are analogous to rotated versions of the surfaces discussed by 
Hartman, Maldacena \cite{Hartman:2013qma} in the eternal $AdS$ 
black hole. It turns out that these surfaces cannot penetrate into 
the Northern/Southern diamond regions $N/S$ beyond a certain point 
(for $dS_{4}$ and higher dimensions): the turning point has a 
real-valued solution only for certain subregions of $N/S$. The limiting 
surface arises as the subregion at $I^{\pm}$ becomes the whole 
space (this limit was identified erroneously in \cite{Narayan:2017xca}).

These surfaces turn out to have various interesting properties, as we
will explore in this paper. We restrict attention to ``top-bottom
symmetric'' surfaces, stretching between a subregion ${\cal A}\in I^+$
and an equivalent subregion at $I^-$: this in some sense simulates
the bulk inner product\ $\Psi^*_{I^+} {\cal O} \Psi_{I^+}$ in
(\ref{dS4/CFT3}), with $\Psi_{I^-}\equiv\Psi_{I^+}^*$\,.
First we will make more elaborate (sec.~2) the construction of these
extremal surfaces for more general subregions, as well as discuss the
limiting surface in more detail. This construction shows that for
multiple disjoint subregions, mutual information vanishes and 
strong subadditivity is saturated. This is reminiscent of finite
temperature systems in $AdS/CFT$, and is perhaps consistent with the
fact that the bulk de Sitter spacetime has a temperature.  We then
argue that there is an effective codim-1 envelope surface formed from
the union or envelope of all the codim-2 surfaces. This leads to
analogs of the entanglement wedge (sec.~3) and a version of subregion
duality, adapting to this de Sitter case the various arguments on the
entanglement wedge in $AdS/CFT$
\cite{Czech:2012bh,Wall:2012uf,Headrick:2014cta}. We close with a
Discussion (sec.~4).

\section{de Sitter space and future-past extremal surfaces}

In $AdS$, surfaces starting at the boundary dip into the radial
direction and exhibit turning points where they begin to return to the
boundary. In $dS$, the boundary at $I^+$ is spatial: surfaces dip into
the time direction (which is holographic) giving a crucial minus sign
that ensures that there is no \emph{real} turning point where the
surface starting at $I^+$ begins to turn back towards $I^+$
\cite{Narayan:2015vda,Narayan:2015oka}. For instance, in the Poincare
slicing\ $ds^2={R_{dS}^2\over\tau^2}(-d\tau^2+dx_i^2)$\,, a strip
subsystem on some boundary Euclidean time $w=const$ slice of $I^+$
with width along $x$ gives a bulk extremal surface $x(\tau)$ described
by\ ${\dot x}^2\equiv ({dx\over d\tau})^2 = {B^2\tau^{2d-2}\over
  1+B^2\tau^{2d-2}}$\ ($B^2>0$). $w$, $x$ can be any of the $x_i$ (no
boundary Euclidean time slice is special). A turning point where the
surface starting at $I^+$ begins to turn back requires $|{\dot
  x}|\ra\infty$ while here $|{\dot x}|\leq 1$.  If such a turning
point existed, the extremal surface, initially dipping into the bulk
time direction (so that $|{dw\over d\tau}|<1$), would have to stop
moving in time and hit $|{dw\over d\tau}|\ra\infty$ which is a
spacelike condition. Thus the surface needs to transit from being
timelike to being spacelike and then again timelike: this appears
incompatible with the extremality of the surface, which is taken to be
smooth.\ (There are also complex extremal surfaces with turning
points, amounting to analytic continuation from the $AdS$
Ryu-Takayanagi surfaces
\cite{Narayan:2015vda,Narayan:2015oka,Sato:2015tta,Miyaji:2015yva}:
their interpretation is unclear, the time parameter taking imaginary
time paths, thus lying outside the original de Sitter time
parametrization.)

Since real surfaces starting at the future boundary $I^+$ keep marching
on into the bulk without returning, it is interesting to ask if they
could instead end at the past boundary $I^-$\ \cite{Narayan:2017xca}.
The bulk probability $\Psi_{dS}^*\Psi_{dS}$ suggests two CFT copies:
so such connected extremal surfaces stretching between $I^\pm$ are
perhaps expected. Towards studying this, we recast $dS_{d+1}$ in the
static coordinatization\
$ds^2 = -(1-r^2/l^2) dt^2 + {dr^2\over 1-r^2/l^2} + r^2 d\Omega_{d-1}^2$\ as
\be\label{dSst}
ds^2_{d+1} = {l^2\over\tau^2} \left(-{d\tau^2\over 1-\tau^2} + (1-\tau^2) dw^2
+ d\Omega_{d-1}^2\right) ,   
\ee
where $\tau={l\over r}\ w={t\over l}$.\ 
Now $\tau$ is ``bulk'' time: $0\leq \tau\leq 1$ define the future-past
universes $F/P$ while the Northern/Southern diamond regions $N/S$ have
$1<\tau\leq\infty$ (with $w$ time).\ 
There are horizons at $\tau=1$: their area is\ ${\pi l^2\over G_4}$\,.
The boundary at $\tau\sim 0$ is Euclidean $R_w\times S^{d-1}$.

Since the asymptotic region enjoys rotational invariance in $S^{d-1}$ as
well as $w$-translations, we could pick either some equatorial plane
of the $S^{d-1}$ or a $w=const$ slice as a boundary Euclidean time slice.
The area functional for a codim-2 surface on an $S^{d-1}$ equatorial plane is
\be\label{areaFnEquator}
S=\ l^{d-1} V_{S^{d-2}} \int {d\tau\over\tau^{d-1}}
   \sqrt{{1\over 1-\tau^2} - (1-\tau^2) (w')^2}\ .
\ee
Such codim-2 extremal surfaces are consistent with the scaling
${l^{d-1}\over G_{d+1}}$ of de Sitter entropy. (The $w=const$ slice
turns out to be difficult to analyse in detail, although certain aspects
such as the leading divergence are straightforward to see.)
Extremizing this gives the surface equation and its area as
\be\label{HMsurf}
         {\dot w}^2 \equiv (1-\tau^2)^2 \Big({dw\over d\tau}\Big)^2
         = {B^2\tau^{2d-2}\over 1-\tau^2
           + B^2\tau^{2d-2}}\ , \quad
  S = {2 l^{d-1} V_{S^{d-2}}\over 4G_{d+1}}
\int_\epsilon^{\tau_*} {d\tau\over\tau^{d-1}}\
{1\over \sqrt{1-\tau^2 + B^2\tau^{2d-2}}}\ .
\ee
Here ${\dot w}$ is the $y$-derivative, with
\be\label{tortoise}
y=\int {d\tau\over 1-\tau^2} = {1\over 2} \log \Big|{1+\tau\over 1-\tau}\Big|
\ee
the ``tortoise'' coordinate, useful near the horizons.
The turning point is the ``deepest'' location to which the surface dips
into in the bulk, before turning around: this is given by
\be\label{tau*}
 |{\dot w}|\ra\infty:\qquad\  1-\tau_*^2+B^2\tau_*^{2d-2}\ =\ 0\ .
\ee
With $B^2>0$, real $\tau_*(B^2)$ arises only if $\tau>1$ \ie\ within $N/S$.
For any finite $B^2>0$, we have ${\dot w}\ra 0^+$ near $\tau\ra 0$, with
${\dot w}<1$ for $\tau<1$ (within $F$) and ${\dot w}\ra 1$ as $\tau\ra
1$.  Overall this gives the smooth ``hourglass''-like red curve in
Figure~\ref{fig1} representing the codim-2 extremal surface
stretching from a subregion $\Delta w\times S^{d-2}$ at $I^+$ to an
equivalent one at $I^-$, intersecting the horizons, turning around
smoothly at $\tau_*$ in $N$/$S$. The full extremal surface for the
subregion consists of the left and right portions of the surface. We
will discuss this more elaborately in what follows.


\subsection{Future-past extremal surfaces for general subregions}

\begin{figure}[h] 
\hspace{2pc}
\includegraphics[width=8pc]{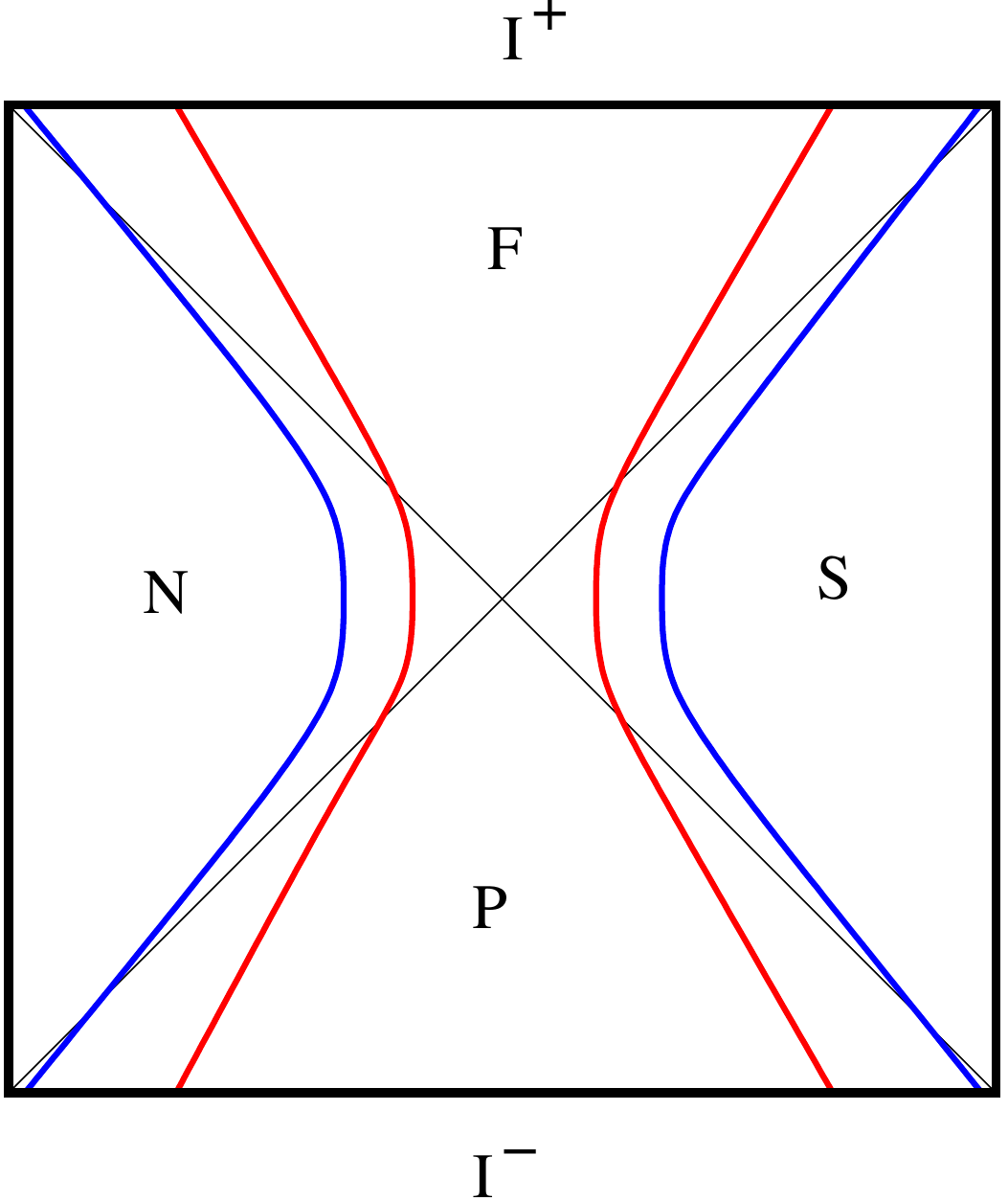}
\hspace{3pc}
\begin{minipage}[b]{24pc}
\caption{{ \label{fig1}
    \footnotesize{Future-past extremal surfaces in de Sitter stretching
      between $I^\pm$. These are akin to rotated Hartman-Maldacena
      \newline surfaces in the eternal $AdS$ black hole. The red curve is
      for \newline generic subregion. The blue curve is a limiting curve
      obtained \newline as the subregion becomes the whole space. \newline }}}
\end{minipage}
\end{figure}

We would like to construct these future-past extremal surfaces for
general subregions at $I^{\pm}$, restricting however to surfaces which
are top-bottom symmetric: as we have stated, this in some sense
simulates the bulk inner product (\ref{dS4/CFT3}). In other words, the
surface is symmetric about the $w=0$ slice passing horizontally
through the middle of the Penrose diagram, \ie\ the top half-surface
(anchored at $I^{+}$) is identical pictorially to the bottom
half-surface (anchored at $I^{-}$);\ see Figure~\ref{fig1}.
To describe these explicitly, consider any $S^{d-1}$ equatorial plane 
and a subregion at $I^{+}$ defined as  
\be\label{width}
{\cal A} \equiv  (w_{L,0}, w_{R,0})\ , 
\qquad w_{R,0}>w_{L,0}\ ,
\ee
where $w_{R,0}$ is the ``right'' edge and $w_{L,0}$ is the ``left'' edge 
of the subregion at $I^{+}$. We take the convention that $I^{+}$ is
parametrized by $w$ with the ``left'' end being $w\ra -\infty$ and the
``right'' end being $w\ra\infty$\ (see \eg\ \cite{Spradlin:2001pw}).
Then $I^{-}$ is parametrized with $w\ra\infty$ at the ``left'' end and
$w\ra -\infty$ at the right end\ (the flow of $\del_w$ is reversed).
This dovetails with taking $w_{R,0}>w_{L,0}$.
To be concrete, consider first the subregion in Figure~\ref{fig1},
which is also left-right symmetric (we will consider more general
subregions later).
The full future-past surface is defined by two sets of equations, one for 
the top half-surface (anchored in the future universe) and the other for the 
bottom half-surface (anchored in the past universe),
\bea\label{wLRy}
top:\quad && w_{L}(y) = w_{L,0} - \int_{0}^{y} dy\, {\dot w_{L}}(B_{L})\, ;\qquad
w_{R}(y) = w_{R,0} + \int_{0}^{y} dy\, {\dot w_{R}}(B_{R})\, , \nonumber\\
bottom:\quad && w_{L}(y) = -w_{L,0} + \int_{0}^{y} dy\, {\dot w_{L}}(B_{L})\, ;
\qquad  w_{R}(y) = -w_{R,0} - \int_{0}^{y} dy\, {\dot w_{R}}(B_{R})\, ,\ \ 
\eea
with ${\dot w_{L}}(B_{L})$ and ${\dot w_{R}}(B_{R})$ given by (\ref{HMsurf}), 
taking the positive square root, with parameters $B_{L}$ and $B_{R}$ 
respectively. The top and bottom half-surfaces are reflections of each other 
about the $w=0$ slice: \ie\ the bottom surface is obtained from the top one 
as $w(y)\ra -w(y)$.
The parameters $B_{L}, B_{R}$ are related to the turning points 
(\ref{tau*}) of the surfaces $w_{L}(y)$ and $w_{R}(y)$ as
\be\label{trngpt}
1-\tau_{*,L}^{2}+B_{L}^{2}\tau_{*,L}^{4}=0\ ,\qquad 
1-\tau_{*,R}^{2}+B_{R}^{2}\tau_{*,R}^{4}=0\ ;\qquad
\tau_{*,L,R}={e^{2y_{*,L,R}}+1\over e^{2y_{*,L,R}}-1}\ .
\ee
The turning points lie in the $N/S$ regions as stated earlier, so we have
used the corresponding expressions for $\tau_{*}(y)$ using (\ref{tortoise}). 
The figure is top-bottom symmetric, as are the extremal surfaces as we 
have stated. Thus the turning points lie on the $w=0$ slice: so we have
for the top half-surface (using (\ref{wLRy}))
\be
0 = w_{L}(y_{*,L}) = w_{L,0} - \int_{0}^{y_{*,L}} dy\, {\dot w_{L}}(B_{L}) \ ,
\qquad
0 = w_{R}(y_{*,R}) = w_{R,0} + \int_{0}^{y_{*,R}} dy\, {\dot w_{R}}(B_{R})\ ,
\ee
which are also automatically satisfied for the bottom half-surface. This 
gives a relation between the boundary conditions at $I^{\pm}$, the 
turning points and the parameters $B$,
\be\label{w0LRintB}
w_{L,0} = \int_{0}^{y_{*,L}} dy\, {\dot w_{L}}(B_{L})\ ,\qquad
w_{R,0} = - \int_{0}^{y_{*,R}} dy\, {\dot w_{R}}(B_{R})\ .
\ee
In other words, the boundary condition $w_{L,0}$ at $I^+$ implies a turning 
point at a specific location $\tau_{*,L}$  (and likewise for the right side 
surface): the bottom part of the left surface can join smoothly to the top 
part only if its turning point matches with that of the top part, which in
turn implies that $w_{L,0}$ for the bottom part should match appropriately 
with the top part. This is why $w^{bottom}_{0}=-w^{top}_{0}$, as we have 
taken in defining the top and bottom half-surfaces (\ref{wLRy}).
In the vicinity of the turning point,
we have $w(\tau)\sim \pm\sqrt{\tau_{*}-\tau}$ for the top part of the
surface: this joins smoothly with $w(\tau)\sim \mp\sqrt{\tau_{*}-\tau}$
from the bottom part of the surface. 
It can now be shown that the integral\ $\int_0^{y_*} dy\, {\dot w}$\ in
fact takes negative values: we will see this explicitly in a special
case later, and it can also be checked numerically. Given this, we see
from (\ref{w0LRintB}) that 
\be
w_{L,0} < 0\ , \qquad w_{R,0}>0\ ,
\ee
for the surface of the form (\ref{wLRy}) above: this has the left and
right parts on the left and right halves of $I^{+}$ respectively, as for
the left-right symmetric subregion with the red curves in
Figure~\ref{fig1}.\
As we see from (\ref{wLRy}), such a future-past surface stretches from 
$w_{0}\in I^{+}$ to $-w_{0}\in I^{-}$, passing through the turning point 
at $w=0$. Then the size of a subregion $A$ (\ref{width}) at $I^{+}$ is
\be\label{width1}
\Delta w = w_{R,0} - w_{L,0} \ ,
\ee
where $w_{0}$'s are given by (\ref{w0LRintB}).
For the left-right symmetric subregion in Figure~\ref{fig1}, we have 
\be\label{leftrightSymmSurf}
w^{top}_{R,0}=-w^{top}_{L,0}\ ,\quad w^{bottom}_{R,0}=-w^{bottom}_{L,0} ;
\quad {\rm and} \quad
w^{top}_{R,0}=-w^{bottom}_{R,0}\ , \quad w^{top}_{L,0}=-w^{bottom}_{L,0} ,
\ee
the last two relations following from top-bottom symmetry. Left-right 
symmetry also means\ ${\dot w_{L}}={\dot w_{R}}\equiv {\dot w}$ and
so the size of the subregion becomes
\be\label{width2}
\Delta w = -2\int_{0}^{y_{*}} dy\, {\dot w}\ ,
\ee
where $y_{*}$ refers to the turning point for both left and right parts of 
the surface, which are now symmetric. This subregion size at $I^{\pm}$ 
can be evaluated, using (\ref{HMsurf}), as
\be\label{width3}
\Delta w = -2\int_{0}^{y_{*}} dy\, {B\tau^{2}\over \sqrt{1-\tau^{2}+B^{2}\tau^{4}}} 
= -2\int_{0}^{y_{*}} dy\, {\dot w} 
= -2\int_{0}^{Y} dy\, {\dot w}\ -\ 2\int_{Y}^{y_{*}} dy\, {\dot w}\ ;\quad 
\tau=\tau(y) .
\ee
In the last expression, we have broken up the integral 
into the contribution outside the horizon and that inside the horizon.  
Near the horizon, where $y\ra\infty$, we have introduced a cutoff $Y$
to regulate the calculation. Now note that ${\dot w}\ra 1$ near the horizon 
so the contribution near the horizon can be estimated as\
$\int_{Y} dy + \int^{Y} dy = Y - Y$, and the apparent divergence cancels: 
the near horizon contribution to $\Delta w$ is nonsingular. This can also 
be seen in the $\tau$-coordinates, where we regulate the near horizon 
region with a cutoff $T$ as $T=1-\varepsilon$ for $\tau<1$ and 
$T=1+\varepsilon$ for $\tau<1$:  this gives\
$\int^{1-\varepsilon} {d\tau\over 1-\tau^{2}} 
+ \int_{1+\varepsilon} {d\tau\over 1-\tau^{2}}
\sim -\log\varepsilon - (-\log\varepsilon)$,\ which is smooth.\
For generic subregion, the surface equation $w(\tau)$ has a turning point 
at a single zero of the denominator: then near the turning point, we have\
$w(\tau)\sim \sqrt{\tau_{*}-\tau}$ from the contribution near the turning point
which is finite. Thus for generic subregion, the width does not grow large 
but remains finite: this occurs for generic values of $0\leq B< {1\over 2}$ 
for $dS_{4}$\,.

\begin{figure}[h] 
\hspace{3pc}
\includegraphics[width=8pc]{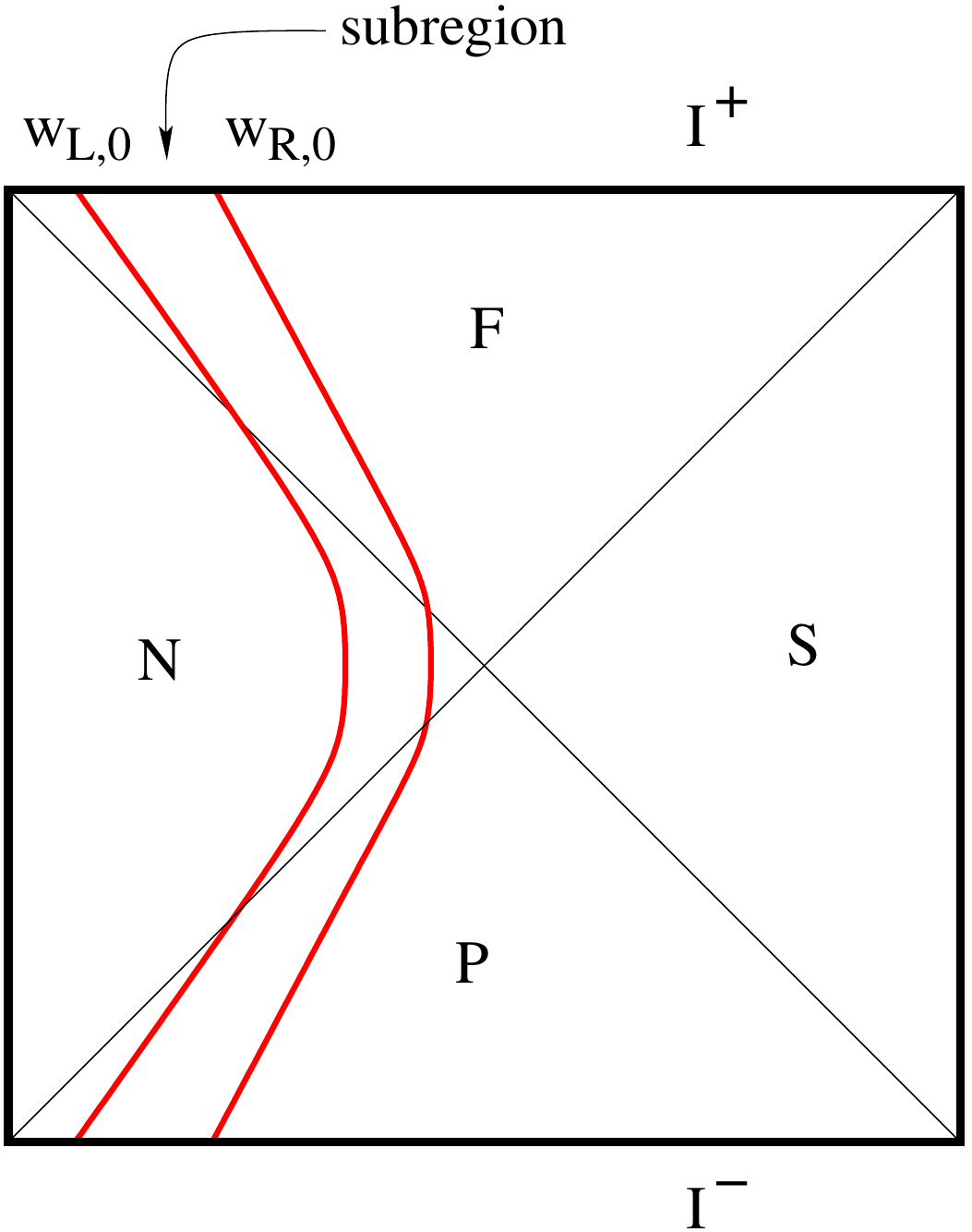}
\hspace{3pc}
\begin{minipage}[b]{24pc}
\caption{{ \label{fig2}
    \footnotesize{Future-past extremal surfaces in de Sitter \newline
      stretching between $I^\pm$ for a generic subregion: these lie \newline
      on some $S^{d-1}$ equatorial plane and have endpoints \newline
      $w_{L,0}$ at the left edge and $w_{R,0}$ at the right 
      edge. \newline\newline }}}
\end{minipage}
\end{figure}
More generally, we see a relation between the equation describing a 
future-past surface and the corresponding boundary condition 
$w_{0}\in I^{+}$,
\be\label{w0tau*}
w(y) = w_{0} \mp \int_{0}^{y} dy\, {\dot w}(B)\qquad \Rightarrow \qquad
w_{0} = \pm \int_{0}^{y_{*}} dy\, {\dot w}(B) \lessgtr 0 \ ,
\ee
so that the top sign corresponds to a surface anchored in the left 
part of $I^{+}$ and the bottom sign to the right part of $I^{+}$. So 
consider \eg\ the more general subregion and the red surfaces in
Figure~\ref{fig2}: this is top-bottom symmetric but not
left-right symmetric. Now we have both $w_{L,0}, w_{R,0}<0$\ (with
$w_{R,0}>w_{L,0}$): thus both surface equations $w_{L,R}(y)$ are of the
form (\ref{w0tau*}) with minus signs: explicitly
\be\label{wLRy2}
top:\qquad  w_{L}(y) = w_{L,0} - \int_{0}^{y} dy\, {\dot w_{L}}(B_{L})\, ;\qquad
w_{R}(y) = w_{R,0} - \int_{0}^{y} dy\, {\dot w_{R}}(B_{R})\, , 
\ee
with the bottom half-surfaces given by\
$w_L^b(y)=-w_L(y) ,\ w_R^b(y)=-w_R(y)$. The $B=0$ surface with
${\dot w}=0$ is a vertical line from $I^+$ to $I^-$, passing through
the bifurcation point.


The full surface for a given subregion (\ref{width}) consists of the left 
and right portions of the surface: the area is thus given by the sum of 
the areas 
\be
S[{\cal A}] = S[w_{L,0}] + S[w_{R,0}]
\ee
of the left and right portions of the surface, where each $S[w]$ is
given by the area in (\ref{HMsurf}). Each of the left and right
portions of the surface of course connects the boundaries of the
equivalent subregions at $I^{+}$ and $I^{-}$, but is disconnected from
the other portion. The fact that the two sides of the full surface are
disconnected from each other dovetails with the fact that a surface
anchored at $I^{+}$ does not return to $I^{+}$ but ends at $I^{-}$.
This disconnectedness has interesting consequences as we will see.

\subsection{The limiting surface and its area}\label{sec:LimitingSurf}

We will now see that the turning point relation (\ref{tau*}) shows the 
existence of a limiting surface at some finite $\tau_{*}$\,:  these are shown 
as the blue curves in Figure~\ref{fig1}. These future-past extremal 
surfaces cannot penetrate deeper than a certain location in the 
Northern/Southern diamond regions $N/S$ beyond the horizons: they are 
in some sense ``repelled'' by the poles (the left/right boundaries in $N/S$).
This is analogous to the repulsive nature of the singularity with regard to 
the Hartman-Maldacena surfaces \cite{Hartman:2013qma} in the eternal 
$AdS$ black hole. 

From (\ref{tau*}), we see that as $B\ra 0$, we have $\tau_{*}\ra 1$. On the 
other hand, we see that the expression cannot vanish if $\tau$ is too 
large: \eg\ beyond $\tau_{*}^{2d-2} B^{2} \gtrsim \tau_{*}^{2}$, the two 
terms cannot cancel. This suggests a nontrivial solution around 
$\tau_{*}^{2d-4}\sim {1\over B^{2}}$.  Focussing on $dS_{4}$, we can 
complete squares for $B\ra {1\over 2}$\ giving
\be
[dS_{4}]\quad B\ra {1\over 2}\,:\qquad  1-\tau^{2}+B^2 \tau^{4}\ \ra\ 
\Big(1-{\tau^{2}\over 2}\Big)^{2}\qquad\Rightarrow\qquad \tau_*\ra\sqrt{2}\ ,
\ee
and the expression in (\ref{tau*}) acquires a double zero.
Now the width integral can acquire a logarithmic divergence: the surface 
is almost completely inside the horizon now (within the $N/S$ regions).
Here we can approximate $\tau\gtrsim 1$ and so the subregion width
(\ref{width1}), (\ref{width2}), can be approximated as\
\be
\Delta w \sim -2\int {d\tau\over\sqrt{1-\tau^{2}+B^{2}\tau^{4}}}\Big|_{\tau_{*}}
\ee
mostly arising from the contribution inside the horizon (ignoring finite 
constants). This is further vindicated since as we have seen above, the 
region near the horizon is smooth and the apparently divergent terms 
cancel after regulating. The subregion is becoming the whole space 
now, since $\Delta w\ra\infty$\,: the left and right edges are going to
the boundaries of $I^{\pm}$.

To see all this more explicitly, we evaluate the width at this double
zero location $B={1\over 2}$\,. We have
\be\label{DeltawSqrt2}
\Delta w = -2\int_{0}^{1-\varepsilon} {d\tau\over 1-\tau^{2}}\, 
{\tau^{2}/2\over 1-\tau^{2}/2} - 2\int_{1+\varepsilon}^{\sqrt{2}-\delta} 
{d\tau\over 1-\tau^{2}}\, {\tau^{2}/2\over 1-\tau^{2}/2}\ .
\ee
The $\varepsilon$ is related to the cutoff $Y$ mentioned earlier near the 
horizon\ (we are symmetrically ``point-splitting'' the horizon), while $\delta$ 
is a cutoff near the turning point $\tau_{*}=\sqrt{2}$ which will illustrate the 
logarithmic divergence. Noting\ 
${1\over 1-\tau^{2}}\, {\tau^{2}/2\over 1-\tau^{2}/2} 
= {1\over 1-\tau^{2}} - {2\over 2-\tau^{2}}$\,, the integrals over $\tau<1$
and over $\tau>1$ are evaluated as
\bea
&& \tau<1:\qquad   \int {d\tau\over 1-\tau^{2}}\, 
{\tau^{2}/2\over 1-\tau^{2}/2} = {1\over 2} \log {1+\tau\over 1-\tau} +
{1\over\sqrt{2}} \log {\sqrt{2}-\tau\over \sqrt{2}+\tau}\ ,\nonumber\\
&& \tau>1:\qquad -\int {d\tau\over \tau^{2}-1}\, {\tau^{2}/2\over 1-\tau^{2}/2} 
=  {1\over 2} \log {1+\tau\over \tau-1} +
{1\over\sqrt{2}} \log {\sqrt{2}-\tau\over \sqrt{2}+\tau}\ .
\eea
Then evaluating (\ref{DeltawSqrt2}) vindicates the statements after
(\ref{width3}) on the near horizon cutoff, giving
\bea\label{DeltawLarge}
\Delta w &=& -\left( \log {1\over\varepsilon} + \log 2 
+ \sqrt{2}\log {\sqrt{2}-1\over\sqrt{2}+1} \right)
- \left( \log {\sqrt{2}+1\over\sqrt{2}-1} + \sqrt{2} \log {\delta\over 2\sqrt{2}} 
- \log {1\over\varepsilon} \right) \nonumber\\
&\sim&  \sqrt{2} \log{1\over\delta}\ .
\eea
Since $\delta\ra 0$, we see that the width $\Delta w$ diverges 
logarithmically at the double zero location $B\ra {1\over 2}$\,, where the 
subregion becomes the whole space. The $\Delta w$ here has identical
contributions from the left and right portions of the surface.

We can also evaluate the area: this again is the sum of the areas of the
left and right portions of the surface, both of which give identical
contributions. For either of them, firstly the divergent part of the area
arises from the region near the boundary, which can be approximated as
the portion outside the horizon,
\be\label{HMareadiv}
S^{{div}} \sim {2 l^{d-1} V_{S^{d-2}}\over 4G_{d+1}}
\int_\epsilon^1 {d\tau\over\tau^{d-1}}\, {1\over \sqrt{1-\tau^2}}\quad
\xrightarrow{\ dS_4\ }\quad {\pi l^2\over G_4} {1\over\epsilon}\ .
\ee
This area law divergence arises for any subregion. 
The scaling of the finite part of the area can be obtained in the limit we 
have discussed above of the subregion becoming the whole space, 
\ie\ $\Delta w\ra\infty$. We have (regulating near the turning point as 
above)
\be\label{areaFin}
S^{fin}\ \sim\ {\pi l^{2}\over G_{4}} \int_{1}^{\sqrt{2}-\delta}
{d\tau\over 1-\tau^{2}/2} 
\ \sim\ {\pi l^{2}\over G_{4}}\, \log {1\over\delta} 
\ \sim\ {\pi l^{2}\over G_{4}}\, c\, \Delta w\ ,
\ee
using the scaling expression (\ref{DeltawLarge}) above (and $c$ is
some $O(1)$ number). 
Thus we see that the finite cutoff-independent part of the area scales 
linearly with the subregion size, as the subregion becomes the whole 
space. This is in some ways related to the linear growth in time of 
entanglement \cite{Hartman:2013qma} in the eternal $AdS$ black hole.

We have seen that this limiting extremal surface arises as the
subregion becomes the whole space: there is an accumulation of
surfaces in the vicinity of this limiting surface, beyond which, these
extremal surfaces do not penetrate.

These surfaces are on $S^{d-1}$ equatorial plane slices: in the $w=const$
slice, similar surfaces appear difficult to identify, although the area law 
divergence is straightforward to see.

\bigskip

\noindent {\bf\emph{Geodesics, or the $dS_{3}$ case}}

$dS_3$ turns out to be special, specially in regard to the above
discussion on limiting surfaces. This case is technically equivalent
to discussing geodesics in any $dS_{d+1}$, and we discuss this now.
The action for timelike geodesics is\
\be
S = \int\sqrt{-g_{\tau\tau}d\tau^2 - g_{ww}dw^2} =
\int {d\tau\over\tau} \sqrt{{1\over 1-\tau^2} - (1-\tau^2) (w')^2}
\ee
which is identical to the action for $dS_3$ extremal surfaces. Then
(\ref{HMsurf}) gives
\be
{\dot w}^2 \equiv (1-\tau^2)^2 (w')^2 = {B^2\tau^2\over 1-(1-B^2)\tau^2}\ ,
\ee
and the turning point where ${\dot w}^{2}\ra\infty$ is
\be
\tau_{*} = {1\over\sqrt{1-B^{2}}}\ .
\ee
For real $\tau_{*}$, we see that the parameter $B$ is restricted to the 
range $0\leq B\leq 1$. However in this case we see that $\tau_{*}\ra\infty$ 
as $B\ra 1$ so that there is no limiting surface at finite $\tau_{*}$: 
geodesic curves reach the North/South poles in $N/S$ as the subregion 
becomes the whole space. This also means that $dS_3$ is special, with
no limiting surface structure of the kind discussed above.
The surface equation can be written as
\be
w(y) = w_{0} \pm \int_{0}^{y} dy\, {\dot w}\ ;\qquad w(\tau_{*})=0\quad
\Rightarrow\quad w_{0}=\mp \int_{0}^{y_{*}} dy\, {\dot w}\ .
\ee
These are top-bottom symmetric, with the turning point $\tau_{*}$ 
lying on the $w=0$ slice in the middle. The integration can be done 
explicitly in this case giving
\be
w_{0} = \int_{0}^{1-\varepsilon} {d\tau\over 1-\tau^{2}} 
{B\tau\over\sqrt{1-(1-B^{2})\tau^{2}}} 
-  \int_{1+\varepsilon}^{\tau_{*}} {d\tau\over \tau^{2}-1} 
{B\tau\over\sqrt{1-(1-B^{2})\tau^{2}}} 
\sim\ {1\over 2} \log (1-B^{2})\ .
\ee
We have, as before, broken up the integral into the portion outside 
the horizon and that inside, regulating near the horizon $\tau=1$ with a 
cutoff $\varepsilon$. The near horizon divergences cancel giving the 
scaling with $B$ above. We see that $w_{0}\ra -\infty$ as $B\ra 1$ or
equivalently $\tau_*\ra\infty$ and $w_{0}\ra 0$ as $B\ra 0$ (\ie\
$\tau_*\ra 1$). The area of the surface (\eg\ the left portion in
Figure~\ref{fig1}) becomes
\be
S = {l\over 2G_{3}} \int_{\epsilon}^{\tau_{*}} {d\tau\over\tau} 
{1\over\sqrt{1-(1-B^{2)}\tau^{2}}}
= {l\over 2G_{3}} \log {2\tau_{*}\over\epsilon}\ .
\ee
The cutoff independent part scales as\ $\log\tau_{*}\sim w_{0}$. Since 
the subregion width is $\Delta w=2w_{0}$ for a left-right symmetric 
subregion, the finite part of the area of the corresponding extremal 
surface grows linearly with the subregion size.

\subsection{Multiple subregions and mutual information}\label{sec:MI}

As we have seen, if we consider top-bottom symmetric subregions and 
the corresponding extremal surfaces, then the turning points always lie
on the $w=0$ slice passing through the middle of the Penrose diagram.
Then from (\ref{w0tau*}), we have\ 
$w_{0} = \pm \int_{0}^{y_{*}} dy\, {\dot w}(B) \lessgtr 0$, \ie\ given the 
location $w_{0}$ of the boundary condition at $I^{+}$, there is a 
unique surface with turning point $y_{*}$ on the $w(y_{*})=0$ slice. This 
surface stretches between $w_{0}\in I^{+}$ and $-w_{0}\in I^{-}$ turning 
at $w=0$.

For a single subregion ${\cal A}\in I^{+}$ defined by the boundary conditions 
$A \equiv (w_{1}, w_{2})$, the equivalent subregion ${\bar {\cal A}}\in I^{-}$
is defined by ${\bar A}\equiv (-w_{1},-w_{2})$.  The extremal surface 
comprises the portion stretching from the top left edge $w_{1}\in I^{+}$ 
and that from the top right edge $w_{2}\in I^{+}$. 
Thus the area of the extremal surface arises from both portions as 
stated earlier, giving
\bea\label{areaAw1w2}
&& S[{\cal A}] = S[w_{1}] + S[w_{2}]\ ; \qquad 
S[w_{0}] \equiv {\pi l^{2}\over G_{4}} \int_{0}^{\tau_{*}(w_{0})} 
{d\tau\over\tau^{2}}\, {1\over\sqrt{1-\tau^{2}+B^{2}\tau^{4}}}\ ,\nonumber\\
&& w_{0} = \pm \int_{0}^{y_{*}} dy\, {\dot w}(B)\ ,\qquad 
1-\tau_{*}^{2} + B^{2}\tau_{*}^{4} = 0\ ,
\eea
noting the relations between the boundary condition at $I^{\pm}$, the 
parameter $B$ and the turning point $\tau_{*}$ (equivalently $y_{*}$).
The turning point above, as we have seen, necessarily lies in the $N/S$ 
regions, satisfying $\tau_{*}\geq 1$: the real, positive, solution to
the quartic with a smooth $B\ra 0$ limit (as $\tau_*\ra 1$) is
\be
\tau_{*} = {\sqrt{1-\sqrt{1-4B^{2}}\over 2B^{2}}}\ .
\ee

\begin{figure}[h] 
\hspace{3pc}
\includegraphics[width=8pc]{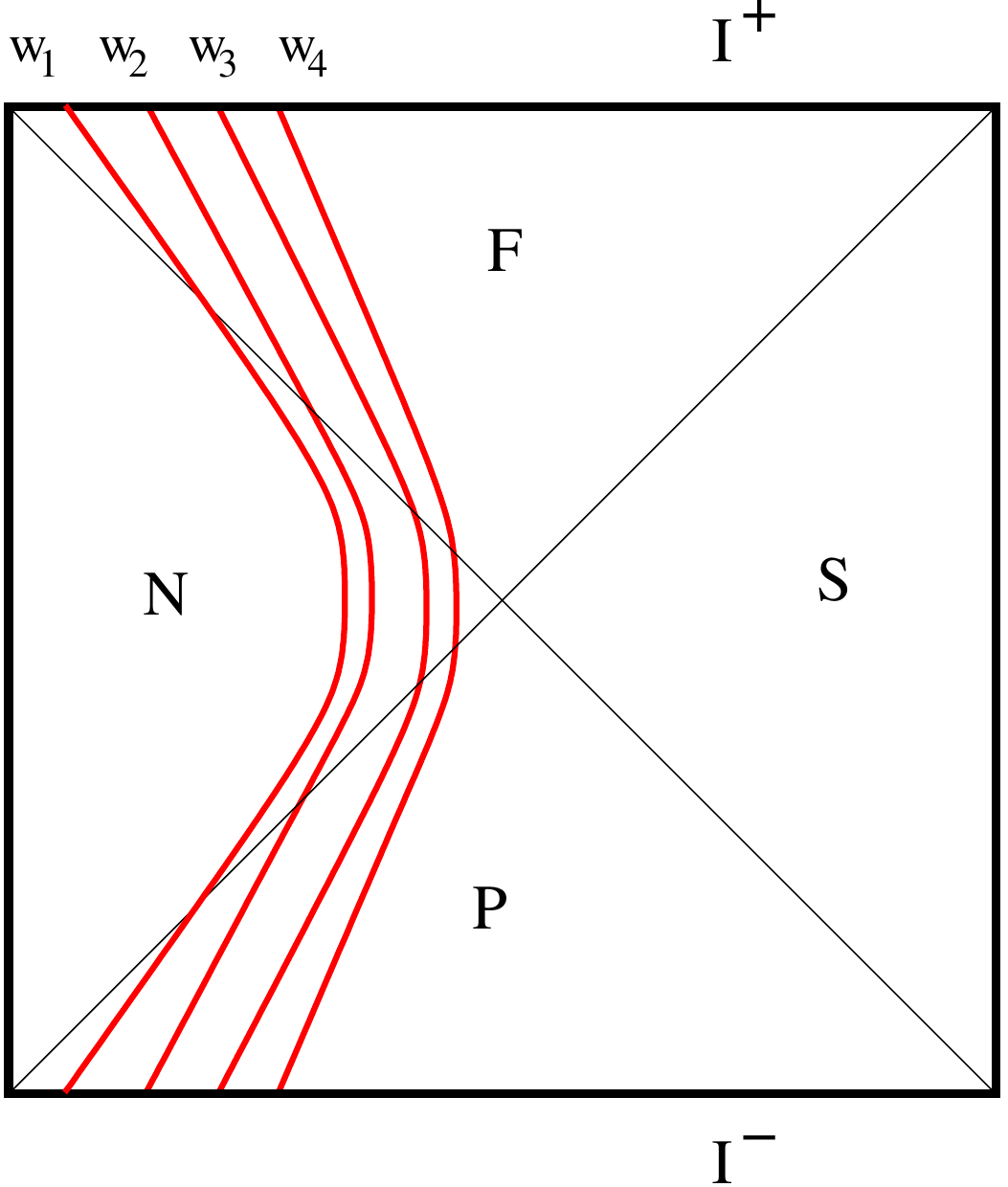}
\hspace{3pc}
\begin{minipage}[b]{22pc}
\caption{{ \label{fig3}
    \footnotesize{Two disjoint subregions $A\equiv (w_{1},w_{2})$ and
      \newline $B\equiv (w_{3},w_{4})$ at $I^{+}$ alongwith the 
      equivalent ones  at $I^{-}$, \newline and the corresponding 
      future-past extremal surfaces.  \newline\newline }}}
\end{minipage}
\end{figure}

This structure of these top-bottom symmetric extremal surfaces and
their areas is special.  Consider two disjoint subregions (restricting
to top-bottom symmetric ones), defined by the boundary conditions at
$I^{+}$,
\be
{\cal A}\equiv (w_{1}, w_{2})\ ,\qquad {\cal B}\equiv (w_{3}, w_{4})\ .
\ee
This is shown in Figure~\ref{fig3}. 
One can then ask if there is any analog of the connected surfaces that 
arise in holographic mutual information defined as
\be
I[{\cal A}, {\cal B}] = S[{\cal A}] + S[{\cal B}] - S[{\cal A}\cup {\cal B}]\ .
\ee
In \eg\ $AdS$, if the subregions ${\cal A}, {\cal B}$ are far apart,
then the minimal area surface for ${\cal A}\cup {\cal B}$ is simply
the two disconnected surfaces around ${\cal A}$ and ${\cal B}$ so
$I[{\cal A}, {\cal B}]$ vanishes. However when $A, B$ are nearby,
there is a new connected surface with lower area \cite{Headrick:2010zt}:
with $A\equiv (x_{1},x_{2})$ and $B\equiv (x_{3},x_{4})$, this new
surface has one part with endpoints $(x_{1},x_{4})$ and another
stretching between $(x_{2},x_{3})$. Since $S[{\cal A}\cup {\cal B}]$
has lower area, we have $I[{\cal A}, {\cal B}]>0$. As the subsystem
separation increases, the area of this new connected surface
increases: at some critical separation, $S[{\cal A}\cup {\cal B}]$
equals that of the two disconnected surfaces and then $I[{\cal A},
  {\cal B}]$ vanishes. This is a large $N$ disentangling transition
for mutual information.

In the present $dS$ case however, it appears that such new surfaces 
connecting the subregions ${\cal A}, {\cal B}$ above cannot exist. If
they did, there 
would be a portion of the surface that stretches between $w_{2}$ and 
$w_{3}$ connecting the adjacent edges of ${\cal A}, {\cal B}$. However this 
requires a turning point for this portion of the surface which starts at 
$w_{2} \in I^{+}$ and returns to $w_{3} \in I^{+}$: as we have seen, 
such turning points do not exist, except the ones we have discussed 
(lying in the $N/S$ regions, which give the future-past surfaces). In other 
words, the only surfaces for a subregion $(w_{2},w_{3})$ comprises 
the future-past surfaces at $w_{2}$ and $w_{3}$, disconnected from 
each other: the left and right components of this are already contained 
in the surfaces for ${\cal A}, {\cal B}$ separately.\ 
Thus it appears that mutual information vanishes always. 
Equivalently, for each of the boundary points $w_{I}$, there is a unique 
future-past surface, so the area is
\be
S[{\cal A}\cup {\cal B}] = S[w_{1}] + S[w_{2}] + S[w_{3}] + S[w_{4}]
= S[{\cal A}] + S[{\cal B}]\ ,
\ee
and $I[{\cal A}, {\cal B}]$ vanishes. The ${\cal A}\cup {\cal B}$
surface is identical to the two disconnected surfaces for ${\cal A}$
and ${\cal B}$ separately.  This is true for any disjoint subregions:
there is no critical separation betweeen the subregions, unlike the
$AdS$ case.

Now consider three adjacent regions ${\cal A}, {\cal B}, {\cal C}$, that
do not overlap: in Figure~\ref{fig3}, the subregions at $I^+$ are
\be
   {\cal A}\equiv (w_1, w_2)\ ,\quad {\cal B}\equiv (w_2, w_3)\ ,\quad
   {\cal C}\equiv (w_3, w_4)\ ,
\ee
so
\be
   {\cal A}\cup {\cal B} \equiv (w_1, w_3)\ ,\quad
   {\cal B}\cup {\cal C} \equiv (w_2, w_4)\ ,\quad
{\cal A}\cup {\cal B}\cup {\cal C} \equiv (w_1, w_4)\ .
\ee
Then using (\ref{areaAw1w2}), the areas give
\be\label{SSA}
S[{\cal A}\cup {\cal B}] + S[{\cal B}\cup {\cal C}]
= S[w_1] + S[w_3] + S[w_2] + S[w_4] 
= S[{\cal A}\cup {\cal B}\cup {\cal C}] + S[{\cal B}]\ .
\ee
Thus strong subadditivity is always saturated, dovetailing in a sense
with vanishing mutual information (and suggesting that tripartite
information \cite{Hayden:2011ag} as well as the entanglement wedge
cross-section \cite{Takayanagi:2017knl} also vanish).  This fact is
independent of the detailed scaling of the area with the subregion
size: it follows from the fact that there is a unique surface
stretching from a given boundary location $w_{0}\in I^{+}$ and that
these are future-past surfaces stretching from $I^{+}\ra I^{-}$, with
no $I^{+}\ra I^{+}$ turning point. The top-bottom symmetry that we have
been focussing on has been a crucial ingredient here.

The above observations might suggest that the surfaces encode nothing,
with no correlations between any two subregions at $I^{+}$. However
the vanishing of mutual information here is in fact reminiscent of a
finite temperature system: for subsystem sizes and separation above
the scale set by the temperature, the entanglement entropy scales
linearly ensuring that mutual information vanishes (see
\eg\ \cite{Fischler:2012uv} for a study of holographic mutual
information in finite temperature $AdS/CFT$). This appears consistent
with the fact that the bulk de Sitter space is a thermodynamic object,
with an entropy and a temperature (in a sense like the $AdS$ black
hole): however it is perhaps a special feature that this vanishing of
mutual information holds for any disjoint subregions, independent of
the separation (in some sense, subregions here are already
``well-separated'' unlike the $AdS$ case).
This is in the leading gravity (large $N$) approximation, with
possibly nonvanishing subleading ($O(1)$) contributions.

\vspace{-1.1mm}

\section{``Entanglement wedge'': future-past surfaces}

We would like to discuss the analogs of the entanglement wedge in
$AdS/CFT$ \cite{Czech:2012bh,Wall:2012uf,Headrick:2014cta} (see also
the reviews \eg\ \cite{Rangamani:2016dms,Harlow:2018fse,Headrick:2019eth})
in the present de Sitter case. Although the present case has very
different features, there are analogs here of the ``entanglement wedge''
and subregion duality.

\subsection{A codim-1 ``envelope'' surface from codim-2 surfaces}\label{sec:EnvlpSurf}

Given some boundary Euclidean time slice and a boundary subregion on
it, we can use de Sitter isometries to generate other equivalent
(codim-1) boundary subregions, since no such slice is special: \eg\ a
subregion with width $\Delta w$ on any $S^{d-1}$ equatorial plane can,
by an asymptotic rotation, be transformed to another subregion on any
$S^{d-1}$ equatorial plane.  This suggests that the true bulk
subregion is a codim-0 subregion with any slice giving a codim-1
subregion. Likewise in the bulk, the codim-2 extremal surface anchored
at the boundary of any codim-1 subregion can be rotated to any
equivalent extremal surface anchored on an equivalent subregion: this
suggests that the natural bulk object is a codim-1 extremal surface
obtained as a ``union'' or ``envelope'' of the family of codim-2
extremal surfaces. Although the only role of the codim-1 envelope
surface is to encode the codim-2 surfaces in its slices, it is the
natural object here since no boundary Euclidean time direction is
sacrosanct. This codim-1 surface will play essential roles in what
follows.

Note that this is not equivalent to defining the boundary subregion as 
the codim-0 boundary subregion $\Delta w\times S^{d-1}$ directly and 
constructing the corresponding codim-1 bulk extremal surface. Such a 
codim-1 extremal surface  in $dS_{d+1}$ is described by the area 
functional\  $S = l^{d}V_{S^{d-1}} \int {d\tau\over \tau^{d}}\
\sqrt{{1\over 1-\tau^2} - (1-\tau^2) (w')^2}$\ and can be seen to scale
as $l^d$. These are structurally similar to codim-2 surfaces in
$dS_{d+2}$, the extremization giving
\be\label{HMcodim1}
 {\dot w}^2 \equiv (1-\tau^2)^2 \Big({dw\over d\tau}\Big)^2
         = {B^2\tau^{2d}\over 1-\tau^2
           + B^2\tau^{2d}}\ , \qquad
 S = {2 l^{d} V_{S^{d-1}}\over 4G_{d+1}}
\int_\epsilon^{\tau_*} {d\tau\over\tau^{d}}\
{1\over \sqrt{1-\tau^2 + B^2\tau^{2d}}}\ ,
\ee
with the turning point
\be
|{\dot w}|\ra\infty:\qquad\
1-\tau_*^2+B^2\tau_*^{2d}\ =\ 0\ .
\ee
These surfaces in $dS_4$ have a leading area divergence\
$S \sim 2 l^{3} V_{S^{2}} \int_\epsilon^{1} {d\tau\over\tau^{3}} 
{1\over\sqrt{1-\tau^2}}= 4\pi l^3 ({1\over\epsilon^2} + \log {2\over\epsilon})$,\ 
thus containing a subleading logarithmic divergence. So these ab initio 
codim-1 surfaces are qualitatively different from the codim-1 ``envelope'' 
surfaces described above. The latter have area scaling as de Sitter 
entropy on each boundary Euclidean slice with no subleading 
logarithmic divergence. We have constructed the codim-2 surfaces
on boundary Euclidean time slices: this crutch appears in accordance
with setting up a subsystem in the dual field theory on some constant
Euclidean time slice and defining entanglement thereof. The effective
codim-1 envelope surface arises as above since no such slice is
special.

\subsection{Domains of dependence and Cauchy horizons}\label{sec:domains}

\noindent {\bf\emph{Domains of dependence:}}\ \ As we have been
saying, the boundary theory is Euclidean, as are the subregions: so
there is no intrinsic boundary time. Boundary Euclidean time
directions, used in the construction of the extremal surfaces above,
are all equivalent: these serve as a crutch to organize the surfaces
simulating entanglement in the dual field theory, but they are simply
spatial directions in the end.


\begin{figure}[h] 
\hspace{4pc}
\begin{minipage}[b]{15pc}
  \includegraphics[width=9pc]{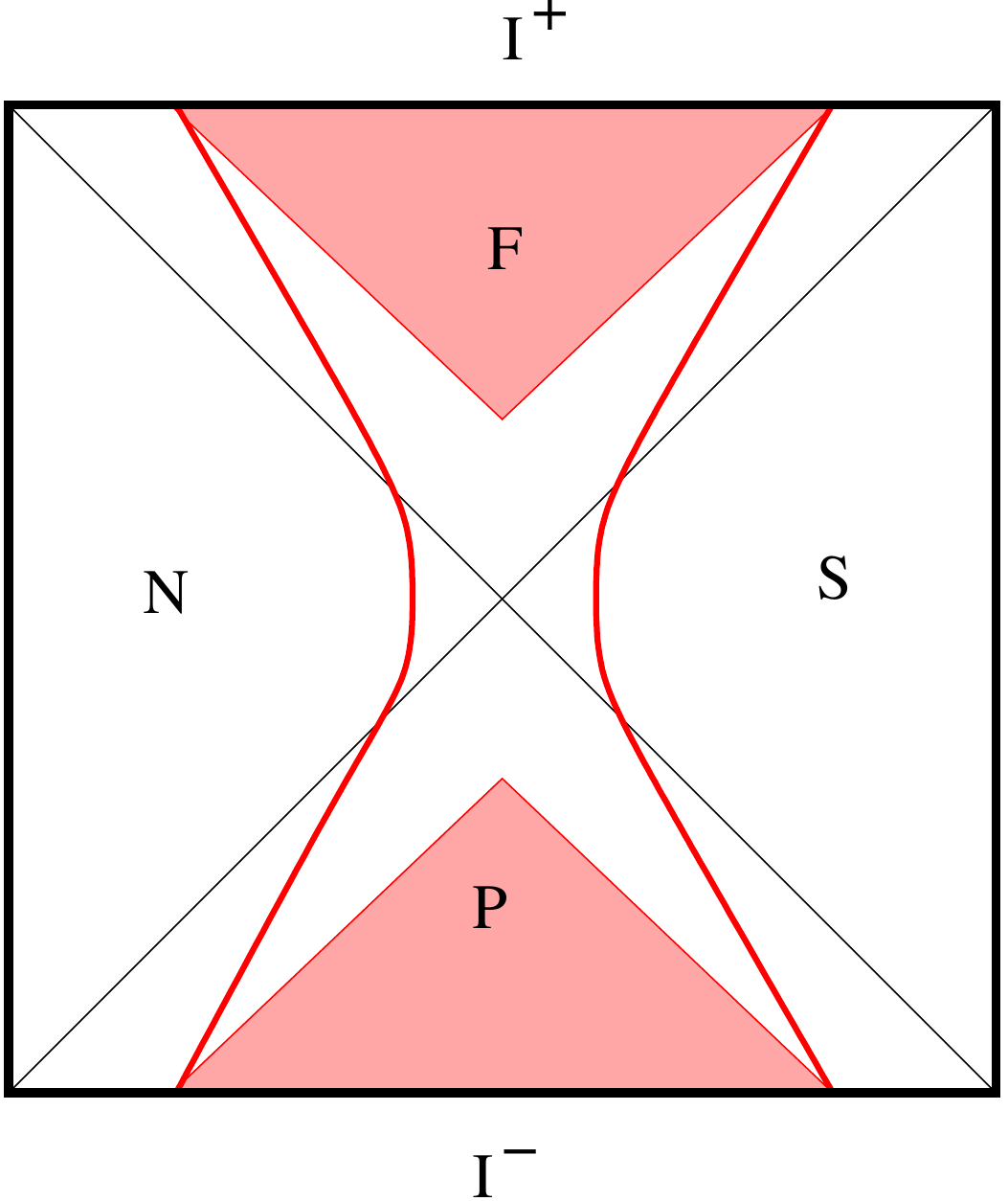} 
  \caption{{\label{fig4} \footnotesize{Generic subregion,
        extremal surface and domain of dependence. } }}
\end{minipage}
\hspace{4pc}
\begin{minipage}[b]{15pc}
\includegraphics[width=9pc]{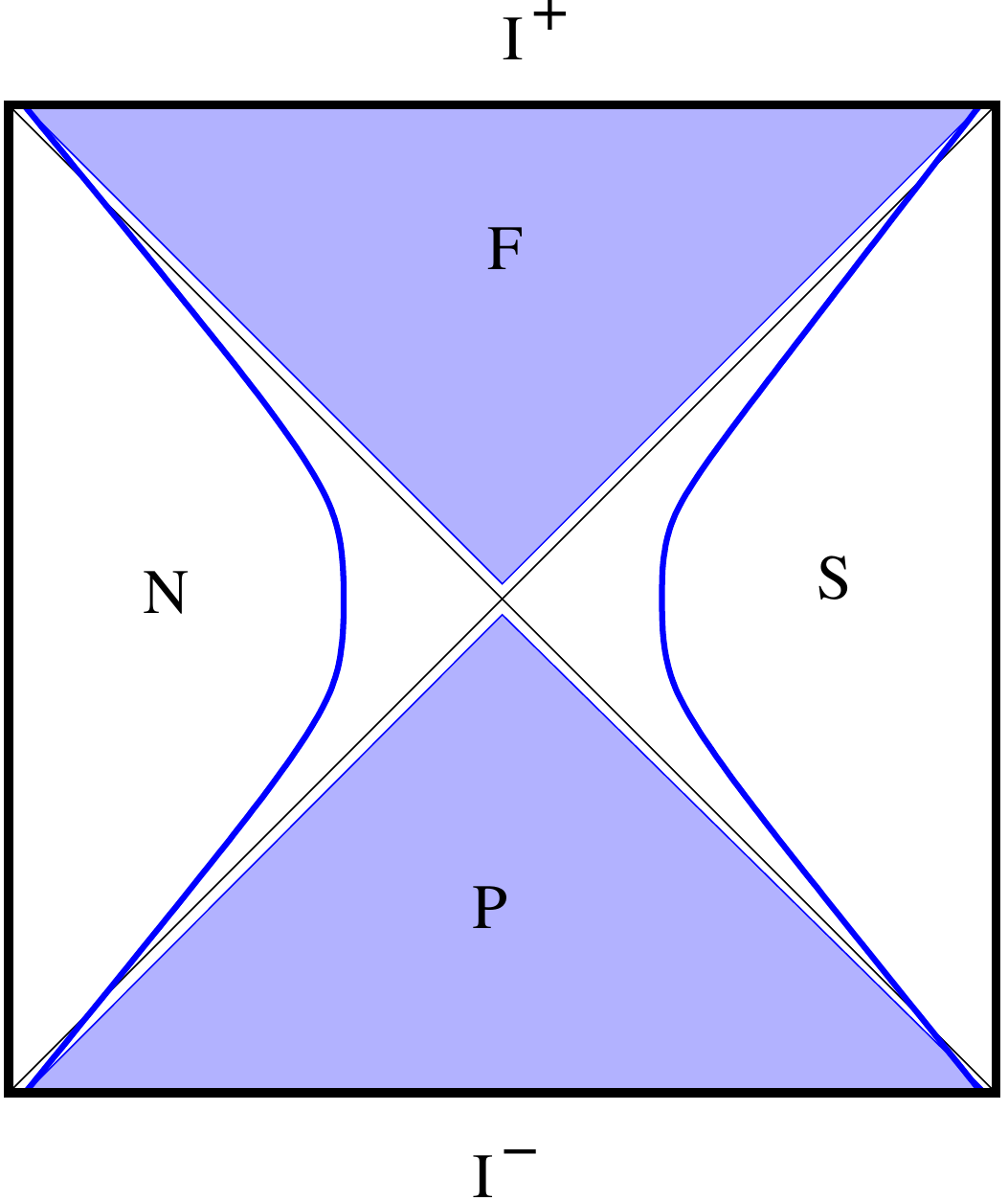} 
\caption{{\label{fig5} \footnotesize{Subregion becoming all $I^+$,
      limiting surface, domain of dependence. }}}
\end{minipage}
\end{figure}
Thus, we take the only natural notion of the domain of dependence to
be that defined from the bulk point of view. The natural subregion in
the Euclidean boundary at $I^+$ is codim-0, \ie\ a 3-dim subspace of
the 3-dim boundary (focussing on $dS_4$). In order that entanglement
be defined on boundary Euclidean time slices of the full space, we
construct this codim-0 subregion ${\cal A}\in I^+$ as the interior of
the boundary at $I^{+}$ of the codim-1 bulk envelope surface we have
discussed above.  In other words, the codim-0 boundary subregion is
the interior of the codim-1 boundary obtained as the ``envelope'' of
all the codim-2 boundaries of the codim-2 subregions. Thus this
codim-0 subregion is essentially the union or envelope of the various
codim-1 subregions (on boundary Euclidean time slices): the latter are
of the form $\Delta w\times S^1$ on some equatorial plane of the $S^2$
in $dS_4$. Thus the codim-0 envelope arising as the union of the
subregions on all such equatorial planes becomes\ $\Delta w\times
S^2$.\ This is represented schematically in Figure~\ref{fig4}
for a subregion symmetrically placed in $I^+$ for convenience for now
(we will discuss more general subregions later). Note that these
Figures are to be regarded as describing the envelope surfaces and the
envelope subregions on some slice (all of which are equivalent).

Given ${\cal A}\in I^+$, its complement ${\cal A}_c$ is the rest of
the boundary.  Then the bulk domain of dependence of ${\cal A}$ is the
past lightcone wedge of $A$, shown by the pink region in $F$ in
Fig.~\ref{fig4}.  Technically, the domain of dependence of
${\cal A}$ is the set of all points $p$ such that any non-spacelike
curve originating at $p$ will necessarily intersect ${\cal A}$ (see
\eg\ \cite{Hawking:1973uf}). In other words, any event occurring at
$p$ will necessarily influence the Cauchy data on ${\cal A}$: it is
clear that this is the past lightcone of ${\cal A}$.
Figure~\ref{fig5} shows the subregion, the corresponding
extremal surface and the domain of dependence in the limit where the
subregion becomes the full space.

Finally, we note again that we are restricting to top-bottom symmetric
surfaces: thus the subregion ${\cal A}\in I^+$ has an equivalent
subregion ${\bar{\cal A}}\in I^-$ (with a future domain of dependence,
its future lightcone wedge).

\medskip

\noindent {\bf\emph{Cauchy horizons:}}\ The boundary of the domain of
dependence $D[{\cal A}]$ is referred to as a Cauchy horizon\ (see
\eg\ \cite{Hawking:1973uf}). This is a boundary for the past Cauchy
development of ${\cal A}\in I^+$\ (or the future Cauchy development of
${\bar{\cal A}}\in I^-$).  From the Cauchy data on ${\cal A}$, it is
not possible to infer events outside $D[{\cal A}]$: in other words, a
point outside $D[{\cal A}]$ could communicate (by sending
particles/light via timelike/null trajectories) to the region outside
${\cal A}$ without influencing ${\cal A}$. Thus the boundary of
$D[{\cal A}]$ is a horizon for past development of the Cauchy data on
${\cal A}$.

For the subregion ${\cal A}$ becoming the whole space, \ie\ ${\cal
  A}\ra I^+$, the past domain of dependence becomes the entire future
universe $F$.  The corresponding Cauchy horizons are then the horizons
bounding $F$: these appear as event horizons to observers in the
static patches which are the Northern/Southern diamonds $N/S$, but
have very different nature for subregions at $I^\pm$.

\medskip

\noindent {\bf\emph{Subregions and ``causal shadows'':}}\ \ We take
the ``causal shadow'' of the subregion ${\cal A}$ to be the region
outside the domain of dependence of ${\cal A}$. We see that this is
the region outside the past lightcone wedge of ${\cal A}$. It is to be
noted that an observer in this causal shadow region can still
communicate with the subregion ${\cal A}$ (via timelike/null
trajectories) so this is not quite like the $AdS$ case where the
causal shadow means no communication exists: however they do not
necessarily influence ${\cal A}$, lying beyond the Cauchy horizons
of ${\cal A}$.

For the boundary subregion taken as ${\cal A}\cup {\cal B}$, the
causal shadow is the region outside $D[{\cal A}]\cup D[{\cal B}]$, the
union of the two domains of dependence.  The fact that we have
decomposed the boundary subregion into two disjoint subregions implies
that the bulk domains of dependence $D[{\cal A}]\cup D[{\cal B}]$ are
distinct from, and smaller than, $D[A\cup B]$.

For the subregion becoming the whole space, \ie\ ${\cal A}\ra I^+$,
the domain of dependence is the entire future universe $F$ and the
causal shadow comprises the entire Northern/Southern diamonds $N\cup
S$.  Similar statements hold for subregions at $I^-$ and the past
universe $P$.

\subsection{Future-past surfaces, ``entanglement wedge'' and 
subregions}

\noindent {\bf\emph{Extremal surfaces and causal shadows:}}\ \ We
continue to focus on top-bottom symmetric subregions and the
corresponding extremal surfaces. So consider a subregion ${\cal A}$ at
the future boundary $I^{+}$ and its equivalent subregion at $I^{-}$
and the area of the corresponding bulk future-past extremal surface
stretching between the two copies of ${\cal A}$ (\eg\ the red curve in
Figure~\ref{fig1}). We would like to interpret the area of this
surface as some sort of entanglement.
Since this procedure requires two copies of the boundary theory, the
entanglement represented by the area of the extremal surface in question
is unlikely to be a boundary quantity, but is likely a bulk one. This is 
analogous to the fact that bulk expectation values are obtained by 
weighting by the bulk probability $|\Psi_{dS}|^2$, along the lines of
$dS/CFT$ (\ref{dS4/CFT3}).
The surfaces can be described as discussed earlier.


Now we note that these future-past extremal surfaces stretching
between $I^\pm$ in fact lie in the ``causal shadow'', as defined
above, of the boundary subregion ${\cal A}$, as we see in
Figure~\ref{fig4}. Firstly note that we are considering the
codim-1 envelope surface obtained from all the codim-2 surfaces, as we
have discussed previously. The extremal surface is anchored at the
boundary of the subregion ${\cal A}$ and is timelike: thus it lies
outside the past lightcone wedge of ${\cal A}$ and so is in the causal
shadow. It appears that this will be true for any Euclidean subregion
${\cal A}$ if the extremal surface is timelike, dipping into the
holographic direction which is time in this case. This would not have
been true if the extremal surface, anchored at $\del {\cal A}$, were
spacelike.  Thus in fact the absence of a turning point where the
surface from $I^+$ begins to return to $I^+$ ensures that the surface
is timelike, thereby lying in the causal shadow of the subregion at
$I^+$ (which is spacelike).

This appears consistent in fact with bulk causality if we take the
area of these future-past extremal surfaces to encode entanglement
pertaining to ${\cal A}$ (independent of the precise nature of the
entanglement). Following the arguments in \cite{Headrick:2014cta},
consider any other spacelike surface ${\cal A}'$, say with wiggles
etc, but with the same boundary $\del {\cal A}'=\del {\cal A}$: thus
${\cal A}'$ is any other subregion homologously equivalent to ${\cal
  A}$. Then ${\cal A}'$ has the same past domain of dependence as the
original subregion ${\cal A}$ and so is expected to be unitarily
equivalent to ${\cal A}$ with regard to entanglement properties: the
two reduced density matrices will likely be related by the unitary
transformation corresponding to bulk time evolution.  Thus we expect
${\cal A}'$ to have the same entanglement properties as ${\cal A}$,
and therefore the same extremal surfaces. So we expect that the
extremal surfaces should lie in the causal shadow since if they did
not, they would contradict the above causality requirement.


\medskip

\noindent {\bf\emph{Extremal surfaces and the ``entanglement wedge'':}}\ \
There have been various interesting investigations on the entanglement 
wedge \cite{Czech:2012bh,Wall:2012uf,Headrick:2014cta} pertaining to 
extremal surfaces in $AdS/CFT$. We will now try to adapt some of 
those ideas and constructs to the present de Sitter case: although the 
spacetime structure here is quite different, we will see that some key 
features have analogs in this case as well.

\begin{figure}[h]
\hspace{5pc}
\includegraphics[width=26pc]{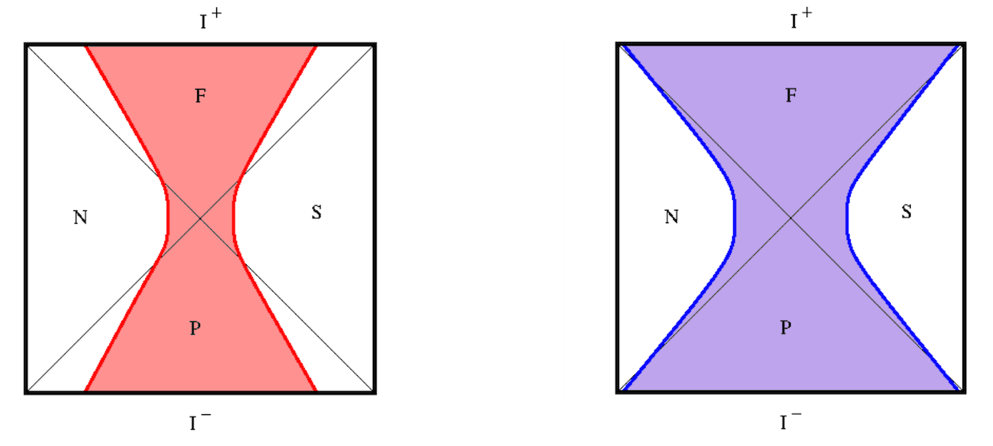} 
\caption{{\label{fig6} \footnotesize{The ``entanglement wedge'' for
generic extremal surface (left), the limiting surface (right).} }}
\end{figure}

\begin{figure}[h]
\hspace{5pc}
  \includegraphics[width=25pc]{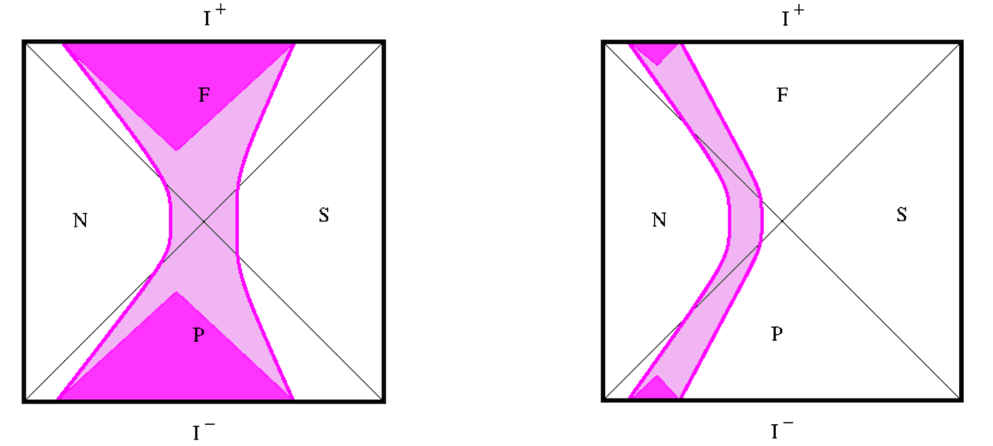} 
\caption{{\label{fig7} \footnotesize{Some more general subregions 
(see \eg\ Figure~\ref{fig2}), the corresponding extremal surfaces 
and the associated ``entanglement wedges''. Also shown within is the bulk 
domain of dependence of these subregions.} }}
\end{figure}

We define the ``entanglement wedge'' here on a constant boundary
Euclidean time slice (\eg\ an $S^{d-1}$ equatorial plane) as the
region enclosed between the extremal surface and the boundary
subregion ${\cal A}$ on $I^\pm$: this gives the shaded regions in
Fig.~\ref{fig6}.  The figure on the left is the entanglement wedge
for the red curve in Fig.~\ref{fig1} while that on the right
shows the limit as the subregion is becoming the whole space. In the
strict limit where the subregion at $I^{\pm}$ is the whole space, the
extremal surface (which is the limiting surface,
sec.~\ref{sec:LimitingSurf}) is essentially contained entirely in the
Northern/Southern diamond regions $N/S$.  We see that the entanglement
wedge now encompasses the future and past universes $F/P$, but is
substantially bigger: it now contains the Cauchy horizons at $\tau=1$
entirely and a substantial portion of the $N/S$ regions.  Thus while
the ``causal wedge'', which can be taken as the past domain of
dependence in this de Sitter case, is necessarily bounded by the
Cauchy horizons, the ``entanglement wedge'' contains more. This is
reminiscent of the fact that in $AdS$, the causal wedge cannot contain
points behind event horizons while the entanglement wedge can.
Since the boundary dual theory is Euclidean here, no boundary Euclidean
time slice can be regarded as sacrosanct as we have been discussing. 
Then the natural subregion is the effective codim-0 subregion on 
$I^{\pm}$ while the bulk region is the interior of the envelope surface 
obtained from the union of the codim-2 extremal surfaces. Thus the true 
bulk entanglement wedge should be taken as the union of the 
entanglement wedge on each of the slices. This generates a codim-0 
bulk region corresponding to the interior of the codim-1 ``envelope'' 
surfaces obtained from the union of the codim-2 surfaces. This is 
represented schematically in Figure~\ref{fig6}\ (see also
Figure~\ref{fig7}).

\medskip

\noindent {\bf\emph{Subregion duality and ``entanglement shadows'':}}\ \ 
The ``entanglement wedge'' above appears to naturally lead to an analog 
of subregion duality in the de Sitter case.

\begin{figure}[h]
\hspace{2pc}
  \includegraphics[width=8pc]{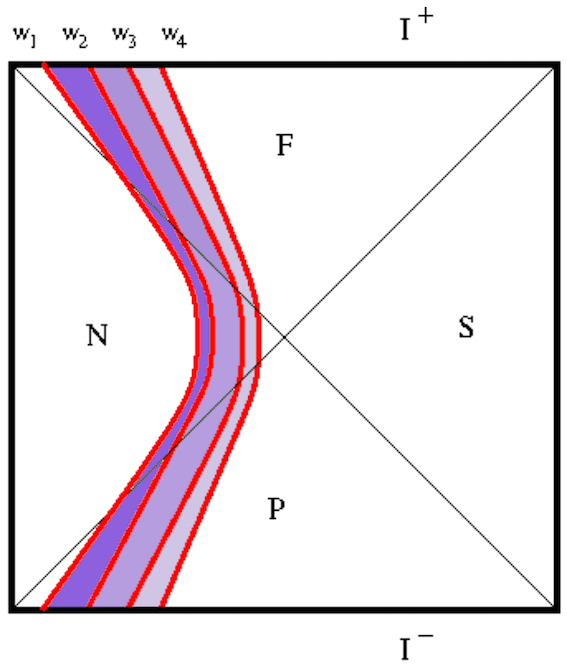}
\hspace{3pc}
\begin{minipage}[b]{24pc}
\caption{{\label{fig8}
\footnotesize{Multiple subregions at $I^{\pm}$ and their 
corresponding \newline top-bottom symmetric future-past extremal surfaces. 
 \newline The corresponding ``entanglement wedges'' suggest an analog 
 \newline of subregion duality. \newline\newline }}}
\end{minipage}
\end{figure}
Given a subregion at $I^+$ and its equivalent subregion at $I^-$, we
have seen that the top-bottom symmetric future-past surfaces define an
``entanglement wedge'' which is the interior of the codim-1 envelope
surface obtained from all the codim-2 extremal surfaces on boundary
Euclidean time slices. For multiple disjoint subregions, the
``entanglement wedges'' thus obtained do not intersect or overlap: the
entanglement wedges foliate the bulk space (see below however). Thus
each boundary subregion at $I^{+}$ (and its equivalent one at $I^{-}$)
lead to a corresponding bulk subregion defined by the ``entanglement
wedge'' of the associated future-past extremal surface.
Figure~\ref{fig8} shows multiple disjoint boundary subregions,
the corresponding top-bottom symmetric future-past extremal surfaces
and the associated ``entanglement wedges''. In some sense, this
appears to be the analog of subregion duality in this de Sitter case
as defined by the future-past extremal surfaces we have been
discussing.

It appears that a large part of the bulk space, including the future
and past universes $F, P$, as well as a large part of the
Northern/Southern diamond regions $N, S$, is obtainable in this manner
as some bulk subregion dual to some boundary subregion. However as we
have seen in sec.~\ref{sec:LimitingSurf}, in $dS_4$ (and higher
dimensions), there is a limiting bulk extremal surface obtained as the
boundary subregion becomes all $I^{\pm}$ : this surface and the
corresponding ``entanglement wedge'' is shown in the right part of
Figure~\ref{fig6}. As is clear, the white (unshaded) regions of the
Northern/Southern diamonds $N/S$, are not accessible by any of these
future-past extremal surfaces. These regions appear to be
``entanglement shadows''. It would be interesting to understand how
precisely these parts of $N/S$ become such shadow regions.

\section{Discussion}

We have discussed various aspects of the future-past extremal surfaces
in de Sitter space, building on previous work \cite{Narayan:2017xca}.
These are rotated analogs of the Hartman-Maldacena surfaces in the
eternal $AdS$ black hole \cite{Hartman:2013qma}. These top-bottom
symmetric codim-2 surfaces (on boundary Euclidean time slices) stretch
between a subregion at $I^+$ in the future universe $F$ and an
equivalent one at $I^-$ in the past universe $P$, with a turning point
in the Northern/Southern diamond regions $N/S$. There exists a
real-valued turning point only for a certain range in $dS_4$ and
higher dimensions: so these surfaces do not penetrate beyond a certain
point in $N/S$, a limiting surface arising as the subregion at $I^\pm$
becomes the whole space (this limit was identified erroneously in
\cite{Narayan:2017xca}). The subregion size $\Delta w$ for this
limiting surface acquires a logarithmic divergence, as we saw in
Sec.~\ref{sec:LimitingSurf}.

For a given subregion at $I^\pm$\ as in Figure~\ref{fig1} or
Figure~\ref{fig2}, the full top-bottom symmetric surface consists of
the left and right portions: each of these connects $I^\pm$ but is
disconnected from the other portion. This disconnectedness reflects
the fact that a surfaces anchored at $I^+$ does not return to $I^+$
but ends at $I^-$. This then implies that for multiple disjoint
subregions, mutual information vanishes, as we discussed in
Sec.~\ref{sec:MI}. This is reminiscent of a finite temperature system
in $AdS/CFT$, perhaps reflecting the fact that the bulk de Sitter
space has a temperature (in some sense, de Sitter space being
thermodynamic is akin to the $AdS$ black hole, rather than pure
$AdS$).  This disconnectedness further implies that strong
subadditivity is saturated. Also, it suggests related points, such as
vanishing tripartite information \cite{Hayden:2011ag} and vanishing
entanglement wedge cross-section \cite{Takayanagi:2017knl}. It is
important to note that the analysis here is all in the leading gravity
(large $N$) approximation: presumably subleading contributions (at
$O(1)$ order) will lead to non-vanishing entanglement quantities. It
would be interesting to explore further properties/inequalities that
these surfaces satisfy, and their implications via $dS/CFT$. In this
regard, since the dual CFT is expected to be non-unitary, it must be
noted that the interpretations thereof (\eg\ vanishing leading mutual
information suggesting no correlations) may be different from those in
ordinary unitary theories.

Since the Euclidean boundary theory has no intrinsic time (all
boundary slices are equivalent), these codim-2 surfaces define an
effective codim-1 ``envelope'' surface and (in $dS_{d+1}$) a codim-0
boundary subregion $\Delta w\times S^{d-1}$, as we saw in
Sec.~\ref{sec:EnvlpSurf}. This leads to analogs of the entanglement
wedge for these future-past extremal surfaces as we saw in Sec.~3.
The associated ``entanglement wedge'' defined as the bulk region
bounded by the extremal surfaces and the boundary subregions is a
bigger region than the ``causal wedge'' (or the domain of dependence)
for the subregion in question. This ``entanglement wedge'' suggests an
analog of subregion duality, Figure~\ref{fig8}: the bulk
subregion enclosed by the extremal surfaces and the boundary
subregions at $I^\pm$ is dual to the boundary subregion in question.
For the limiting surface, corresponding to the boundary subregion
becoming all of $I^+$, the ``entanglement wedge'' covers the
future-past universes $F/P$ as well as a substantial part of the
Northern/Southern diamond regions $N/S$.  However there is a
substantial ``shadow'' part of $N/S$ which is not accessible via these
extremal surfaces for $dS_4$ and higher dimensions\ (it is interesting
to ask if these regions of $N/S$ can be described by \eg\ analogs of
mirror operators \cite{Papadodimas:2013jku}).

Our analysis here has been based on the geometric properties of these
future-past extremal surfaces, and therefore heuristic in some
essential sense. It would be interesting to understand more directly
if these heuristic geometric observations can be made more precise
towards better understanding the physical interpretation, for instance
via the analog in $dS/CFT$ of the interrelations between the
entanglement wedge, modular flow, relative entropy, error correction
codes and so on \cite{Almheiri:2014lwa,Jafferis:2015del,Dong:2016eik}
(see also the reviews
\cite{Rangamani:2016dms,Harlow:2018fse,Headrick:2019eth}). It would
also be interesting to understand the role of HKLL bulk reconstruction
\cite{Hamilton:2005ju,Hamilton:2006az,Hamilton:2006fh} with regard to
these future-past extremal surfaces\ (see \eg\ \cite{Anninos:2017eib}
in the higher spin $dS/CFT$ context; see also
\cite{Xiao:2014uea,Sarkar:2014dma}).

Our discussions so far have focussed on bulk extremal surfaces.
Operationally we have considered a boundary (spatial) subregion at
$I^+$ and its equivalent boundary subregion at $I^-$, and extremal
surfaces stretching between them through the holographic (time)
direction. The top-bottom symmetry in some sense simulates the bulk
inner product $\Psi_{I^+}^* {\cal O} \Psi_{I^+}$, with $\Psi_{I^-}
\equiv \Psi_{I^+}^*$\ (it is in some sense reminiscent of the in-in
formalism).\ By focussing first on boundary Euclidean time slices, we
have simulated setting up entanglement entropy in the dual field
theory, leading to codim-2 surfaces: the fact that these slices are
all equivalent leads to an effective codim-1 envelope surface. The
area of these future-past codim-2 surfaces is positive and exhibits
(for $dS_4$) an area law divergence ${l^2\over G_4} {1\over\epsilon}$
for generic subregions: the finite part scales as\ ${l^2\over G_4}
\Delta w$ linearly with the subregion size $\Delta w$ as the
subregions become all $I^\pm$. The coefficients scale as $dS_4$
entropy, which is akin to the number of degrees of freedom in the dual
$CFT$. The fact that these future-past surfaces are defined with two
boundaries suggests that the area is a sort of bulk entanglement
entropy, especially if we take the (analog of) subregion duality above
seriously. Thus overall these future-past extremal surfaces are
perhaps best interpreted as a way to organize bulk entanglement in
terms of boundary subregions in de Sitter space (in some sense, these
surfaces suggest entanglement between timelike separated regions;
see below). It would be interesting to clarify this further.
Relatedly, it would also be useful to explore more general boundary
subregions (\eg\ caps on the sphere $S^{d-1}$ etc): in such cases we
feel similar future-past extremal surfaces will arise but these appear 
difficult to analyse in detail, so it is unclear if the various
features we have noted here will hold more generally\ (including for
subregions that are not top-bottom symmetric, as might arise under
\eg\ a shock wave perturbation in the bulk).

From the dual point of view, ghost-like CFTs as
\cite{Maldacena:2002vr}, \cite{Anninos:2011ui}, might suggest, are
expected to have negative norm states/configurations, thus suggesting
``negative entanglement''.  Various investigations involving
ghost-like theories, including simple toy quantum-mechanical models of
``ghost-spins'', in fact exhibit this non-positive entanglement quite
explicitly, \eg\ \cite{Narayan:2016xwq,Jatkar:2017jwz}\ (reviewed in
\cite{Narayan:2019pjl}). However, ``correlated'' states entangling
identical ghost-spins between two copies of ghost-spin ensembles can
be shown to have positive norm, reduced density matrices (RDMs) and
entanglement.  Considering two copies of 3-dim $N$-level ghost-spin
systems as microscopic realizations in the universality class of
ghost-$CFT_3$'s dual to $dS_4$ with $N \sim {l^2\over G_4}$\ finite
albeit large, ``correlated'' states \cite{Narayan:2017xca} of the
form\ $|\psi\rangle = \sum \psi^{i_n^F,i_n^P} |i_n\ran_{{}_F}
|i_n\ran_{{}_P}$\ entangling ghost-spin configurations
$|i_n\ran_{{}_F}$ from $CFT_F$ at $I^+$ with \emph{identical} ones
$|i_n\ran_{{}_P}$ from $CFT_P$ at $I^-$, are entirely positive, giving
positive RDM and entanglement. These are in some sense consistent with
the top-bottom symmetric future-past surfaces we have been discussing,
stretching between $I^\pm$. Bulk time evolution maps states at $I^-$
to $I^+$\ \cite{Witten:2001kn}, and suggests the states above are
unitarily equivalent to entangled states in two $CFT_F$ copies solely
at $I^+$.  $CFT_P\equiv CFT_F$ implies a single $CFT$, but the state
above is perhaps best regarded as a particular entangled slice in a
doubled system, akin to the thermofield double dual to the eternal
$AdS$ black hole \cite{Maldacena:2001kr}. This suggests the
speculation \cite{Narayan:2017xca} that $dS_4$ is perhaps
approximately dual to $CFT_F\times CFT_P$ (or $CFT_F\times CFT_F$)
entangled as above, $dS_4$ entropy arising as entanglement. See also
\cite{Arias:2019pzy} for a related discussion.

It would appear that our discussions here are consistent with the
natural holographic screens for de Sitter space (obtained via mapping
along light rays) being at future and past timelike infinity
\cite{Bousso:1999cb}\ (see \cite{wittenStrings98} for an early
discussion of holography and asymptotics, and elaborated on for de
Sitter space \cite{Strominger:2001pn,Witten:2001kn}). Both boundaries
$I^\pm$ are required: these are thus preferred screens for anchoring
the future-past extremal surfaces.  Since these surfaces intersect all
$\tau=const$ surfaces in $F/P$ precisely once, it would appear that
moving the screens towards the interior (\ie\ moving the screens from
$I^\pm$ at $\tau=\epsilon\sim 0$ to say $\tau=\tau_0$ with
$\epsilon<\tau_0\ll 1$) does not affect the construction of these
extremal surfaces. How this gels with \eg\ the area law in
\cite{Bousso:2015mqa} will be interesting to understand. It would also
be interesting to understand other screens such as the Poincare
horizon \cite{Bousso:1999cb}: see \eg\ \cite{Sanches:2016sxy}.  One
might hope that the considerations here may help in understanding and
organizing holography for cosmologies more generally.

Finally, all our explorations here have been analogs of the classical
RT/HRT story in $AdS/CFT$, viewed via a $dS/CFT$ perspective (see
\cite{Lewkowycz:2019xse,Geng:2019ruz} for a different approach, based
on the dS/dS correspondence \cite{Alishahiha:2004md}): it would be
interesting to understand if these investigations can be obtained via
analogs of \cite{Casini:2011kv} or bulk replicas
\cite{Lewkowycz:2013nqa,Dong:2016hjy}.  It would then be interesting
to understand subleading corrections, quantum extremal surfaces
\cite{Faulkner:2013ana,Engelhardt:2014gca} and beyond.

\vspace{16mm}

{\footnotesize \noindent {\bf Acknowledgements:}\ \ It is a pleasure
  to thank Dionysios Anninos, Rajesh Gopakumar, Tom Hartman, Matt
  Headrick, Kedar Kolekar, Alok Laddha, R. Loganayagam and especially
  Suvrat Raju for helpful discussions and comments on a draft. I've
  also benefitted from early discussions with Dionysios Anninos, Juan
  Maldacena, Kyriakos Papadodimas and Shahin Sheikh-Jabbari on
  previous work. This work is partially supported by a grant to CMI
  from the Infosys Foundation.  }

\vspace{3mm}

\end{document}